# Preliminary Design of Integrated Digitizer Base for Photomultiplier Tube

Tao Xue, *Member, IEEE*, Jinfu Zhu, *Student Member, IEEE*, Guanghua Gong, Liangjun Wei, Jianmin Li

***Abstract*–In many physical experiments and applications with photomultiplier tubes (PMTs), high-speed pulse digitalization is used extensively. The current pulse from a PMT has a rising time of several to a hundred nanoseconds. The current pulse shape can be used to obtain the pulse energy and discriminate between different particles. The conventional readout system with a PMT uses the discrete or integrated pulse-shaping circuits and analogue-to-digital converter boards which are based on the NIM or VME system to implement the data acquisition function. In many applications, such as security instruments, there are only several PMTs used in the whole system, or in some distributed applications, long-distance analogue cable with high-voltage power cable is not applicable due to the tremendous number of channels and enormous distributed space. In this paper, an integrated digitizer base is designed and interfaced with a $Cs_2LiYCl_6$ (CLYC) scintillator with a PMT. The CLYC scintillator has the ability to realize the pulse shape discrimination (PSD) between the neutron and gamma-rays. There is only one Gigabit of unscreened twisted-pair cable needed for over 700 Mbps TCP data throughput and more than 10 W power. The boards of high voltage supply, current sensitive preamplifier, 500 MSPS/12-Bit ADC (Analog-to-Digital Converter), readout module based on ZYNQ SoC (System on Chip), power over Ethernet, and user interface circuits are stacked with board-to-board connectors. The system can achieve 4.6% gamma-ray energy resolution at 662 keV with a 20-µs pulse integral and PSD figure-of-merit of ~3.0 for the whole energy of neutrons and gamma-rays. This integrated digitizer base can also be interfaced with other detectors with a PMT.**

***Index Terms*—Integrated Digitizer Base, Photomultiplier Tube, CLYC scintillator, pulse shape discrimination**

## I. Introduction

In the fields of nuclear and particle physics and their applications, the PMT (photomultiplier tube) is extensively used to convert the incident light to the current and multiply it. The current pulse shape from a PMT can be used to discriminate between different particles, such as neutron-gamma discrimination in liquid scintillators [1] and alpha-gamma discrimination in $LaBr_3$:Ce and $LaCl_3$:Ce [2]. At the same time, the pulse energy can be obtained by the current pulse integral.

In recent years, the dual gamma-neutron detector based on the $Cs_2LiYCl_6:Ce^{3+}$ (CLYC) scintillator [3] has attracted much attention and been developed in both scientific research [4-6] and in applications, such as radiation-monitoring devices [7-8] and imaging systems [9]. In this paper, an integrated digitizer base for PMTs is designed and deployed with a CLYC scintillator (it can also be used in other types of detectors with PMTs). This design can realize energy spectroscopy display and pulse shape discrimination simultaneously.

This paper is organized as follows. Section II presents the design of the integrated PMT base, including hardware, firmware, and software design. Section III introduces a moderated $^{252}$Cf source simulation based on Geant4 and experiments with $^{252}$Cf source and different gamma-ray sources. Section IV presents the results of experiments and analyzes the pulse shapes, linearity, energy resolution, and figure-of-merit (FoM). Finally, Section V makes a summary and proposes future work.

## II. The design of an integrated PMT base

### A. Hardware design

In this preliminary design of the integrated digitizer base for PMTs, the 1 GSPS (Giga sample per second) 8-Bit high-speed ADC (Analog-to-Digital Converter) and high-speed current sensitive preamplifier are used for the digitalization of pulse with a few nanoseconds' rising edge [10]. The ZYNQ SoC (system on chip) is used due to its compact footprint, combination of FPGA and ARM processors, relatively low power consumption, and Gigabit Ethernet interface. Only a single category-5 cable is needed for data link (more than 700 Mbps TCP data throughput), synchronized timing (less than 50 ns jitter), and power (more than 10 W power can be delivered). Fig. 1 and Fig. 2 show the architecture [10] and the actual object of this prototype design, respectively. The completely integrated digitizer base consists of a PMT socket, six basic circuit boards, ZYNQ readout module [11], mechanical shell, and supporting studs.

The PMT socket is used to adapt to the PMT, which is R6231-100 from Hamamatsu and coupled with CLYC. The diameter of the six basic circuit boards is 64 mm. The board of high voltage supply, current sensitive preamplifier and main amplifier, ADC subsystem, ZYNQ readout module carrier, PoE power, and user interface are stacked with screw pillars (brass threaded, hex standoff). The cylindrical mechanical shell is made out of aluminum alloy (diameter of 86 mm).

The board of high voltage is designed to interface with the pins of the PMT with a dedicated socket. It provides the bias high voltage -680V through a high-voltage generator module (Maxim 2000 V and 0.5 mA) from XPPOWER.

The authors are with the Key Laboratory of Particle & Radiation Imaging (Tsinghua University), Ministry of Education, Beijing, 100084, China.

The corresponding author is Tao Xue (telephone: 8610-62789104, e-mail: gbe.tao.xue@gmail.com).

Jinfu Zhu is the co-corresponding author, majors in nuclear electronics and instrumentations (e-mail: zhujf16@mails.tsinghua.edu.cn).

The board of current sensitive preamplifier is designed to amplify the current (~μA) from the PMT and match the input scale of ADC. Fig. 3 is the schematic of the PMT equivalent output and current sensitive preamplifier. The current sensitive preamplifier belongs to Trans-Impedance Amplifier (TIA) [12]. The -3 dB bandwidth $f_{-3dB}$ of TIA is calculated as equation (1).

$$f_{-3dB} = \sqrt{\frac{GBP}{2\pi R_F C_S}} \quad (1)$$

*GBP* is the gain bandwidth product of the amplifier (1.6 GHz for OPA657). $R_F$ is the feedback resistor (trans-impedance gain). The feedback capacitor $C_F$ and the source capacitance $C_S$ are used for stability. $C_S$ is the sum of the input capacitance and is calculated in equation (2). $C_D$ is the output capacitance from the PMT. $C_{CM}$ (common-mode input capacitance) and $C_{DIFF}$ (differential-mode input capacitance) depend on the amplifier.

$$C_S = C_D + C_{CM} + C_{DIFF} = 46.3 + 0.7 + 4.5 = 51.5pF \quad (2)$$

The bandwidth of the output signal from the PMT is about 70 MHz by FFT (fast Fourier transform) analysis. It is measured from the anode output of the PMT with an oscilloscope (WaveRunner 8408, coupled with DC 50 ohm).

The $f_{-3dB}$ should be more than 70 MHz by the requirement of the signal from the PMT, so the $R_F$ is limited to be less than 1.373k ohm. In this design, $R_F$ is set to 1k ohm, and $C_F$ is set to 1.8 pF in order to make the TIA operate at high frequency [13]. Fig. 4 shows the PSpice simulation of frequency response of the TIA.

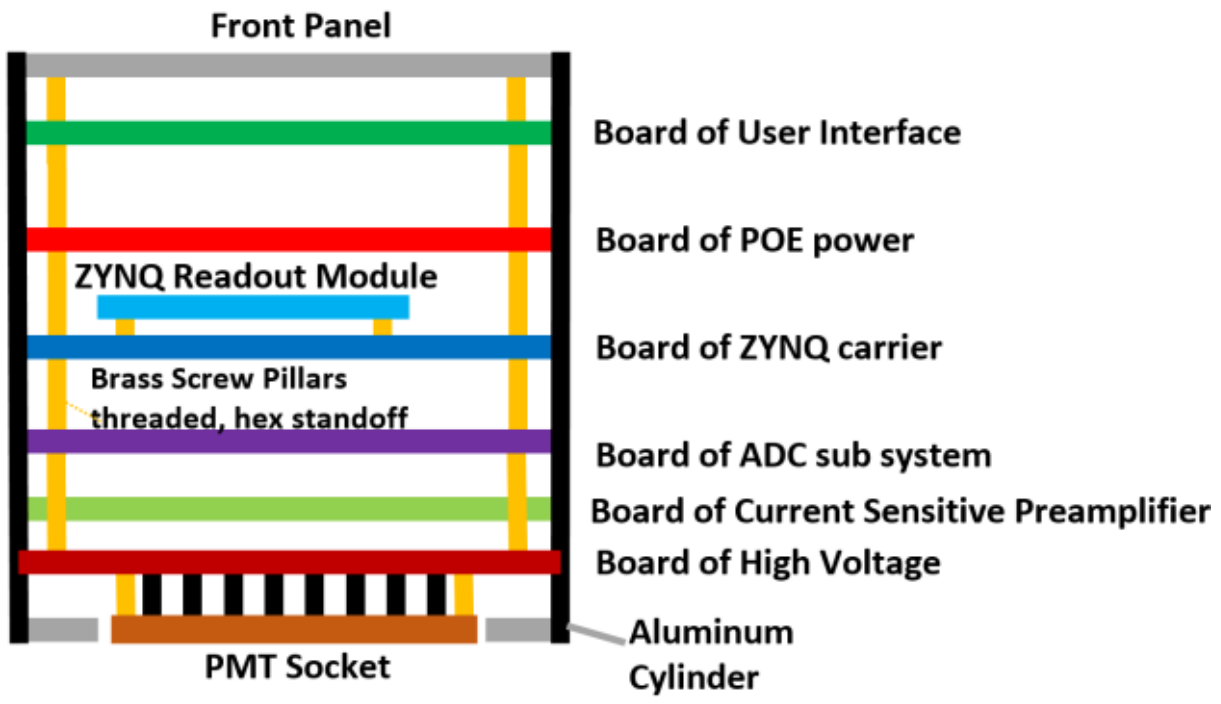

Fig. 1. Architecture of the integrated digitizer base for PMT.

The board of the ADC subsystem contains a 500 MSPS 12-Bit ISLA212P50, ultralow jitter oscillator CCHD-575, voltage-controlled oscillator (VCO) ADF4360-7, differential amplifier LHM6554 for ADC input driver, digital-to-analogue converter (DAC) AD5686 for input bias adjustment, and the associated low-noise power supply. The ENOB (effective number of bits) of the ADC is 10.7 at 30 MHz measured by a standard sine wave according to IEEE 1241-2000 standard.

The board of PoE power management contains the PoE powered device (PD) controller Si3402. The maximum power allowed to supply by the PoE board is more than 10 W according to IEEE 802.3af standard. Fig. 5 describes the efficiency of the PoE PD and its output ripple. The PoE PD efficiency is ~80% when the output power is more than 4 W, and the ripple is less than 35 mV. According to our test, the total power consumption of the integrated digital PMT base is ~5.6 W. Table I shows the distribution of power consumption.

TABLE I
THE DISTRIBUTION OF POWER CONSUMPTION

| Board | Power consumption (W) |
|---|---|
| High voltage supply | 0.3 |
| Current sensitive preamplifier and main amplifier | 0.4 |
| ADC subsystem | 1.6 |
| ZYNQ readout module | 1.8 |
| PoE power | 1.1 |
| User interface | 0.4 |
| Total power consumption | 5.6 |

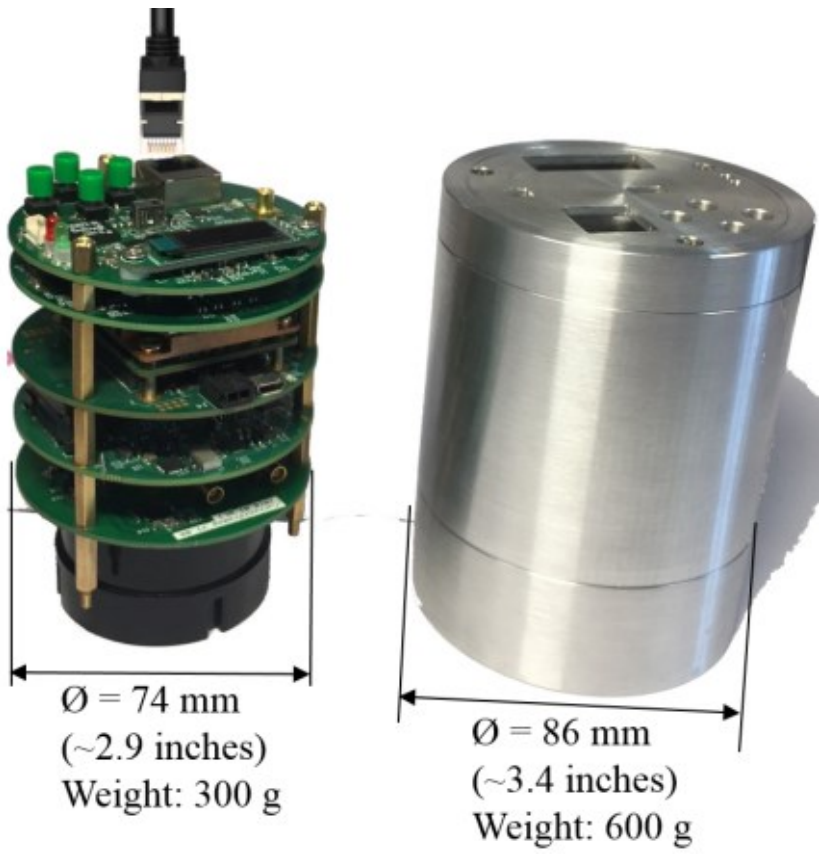

Fig. 2. Actual object of the integrated digitizer base.

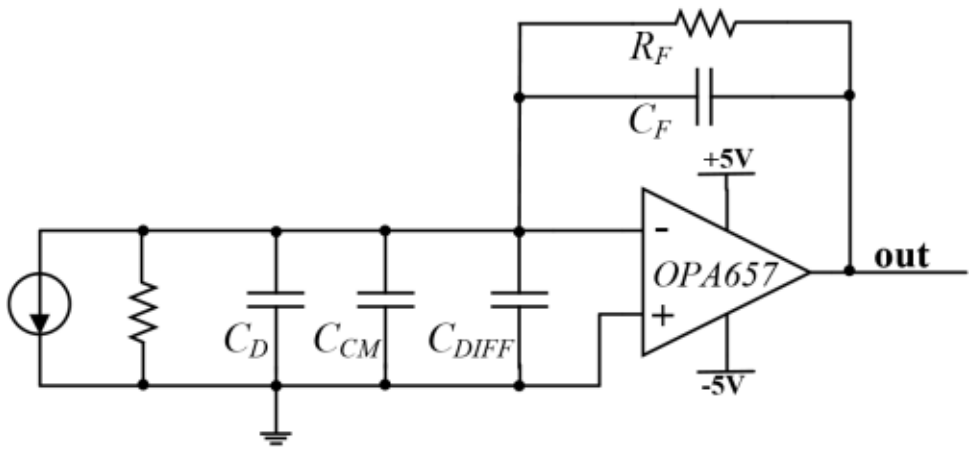

Fig. 3. The circuit of PMT equivalent output and current sensitive preamplifier.

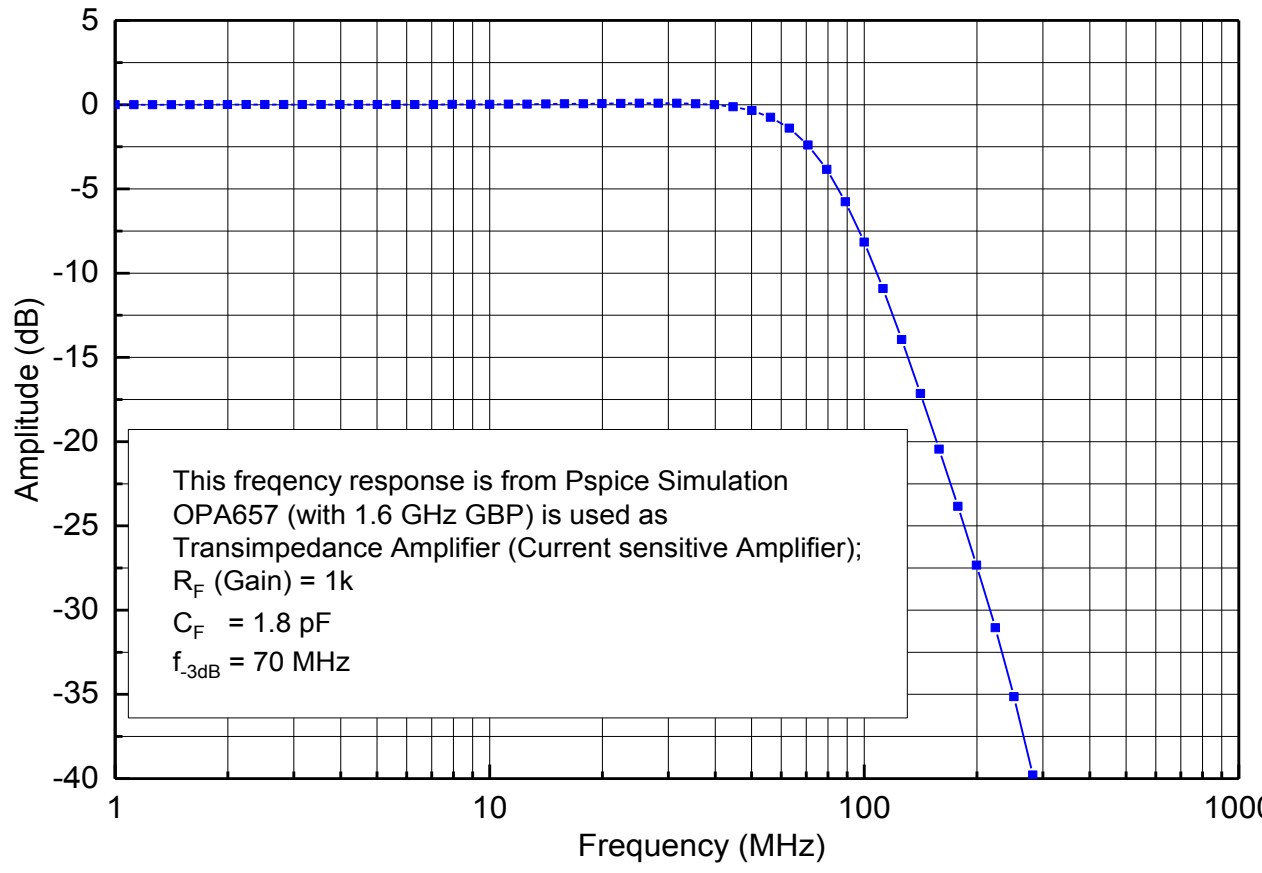

Fig. 4. Frequency response of TIA from PSpice simulation.

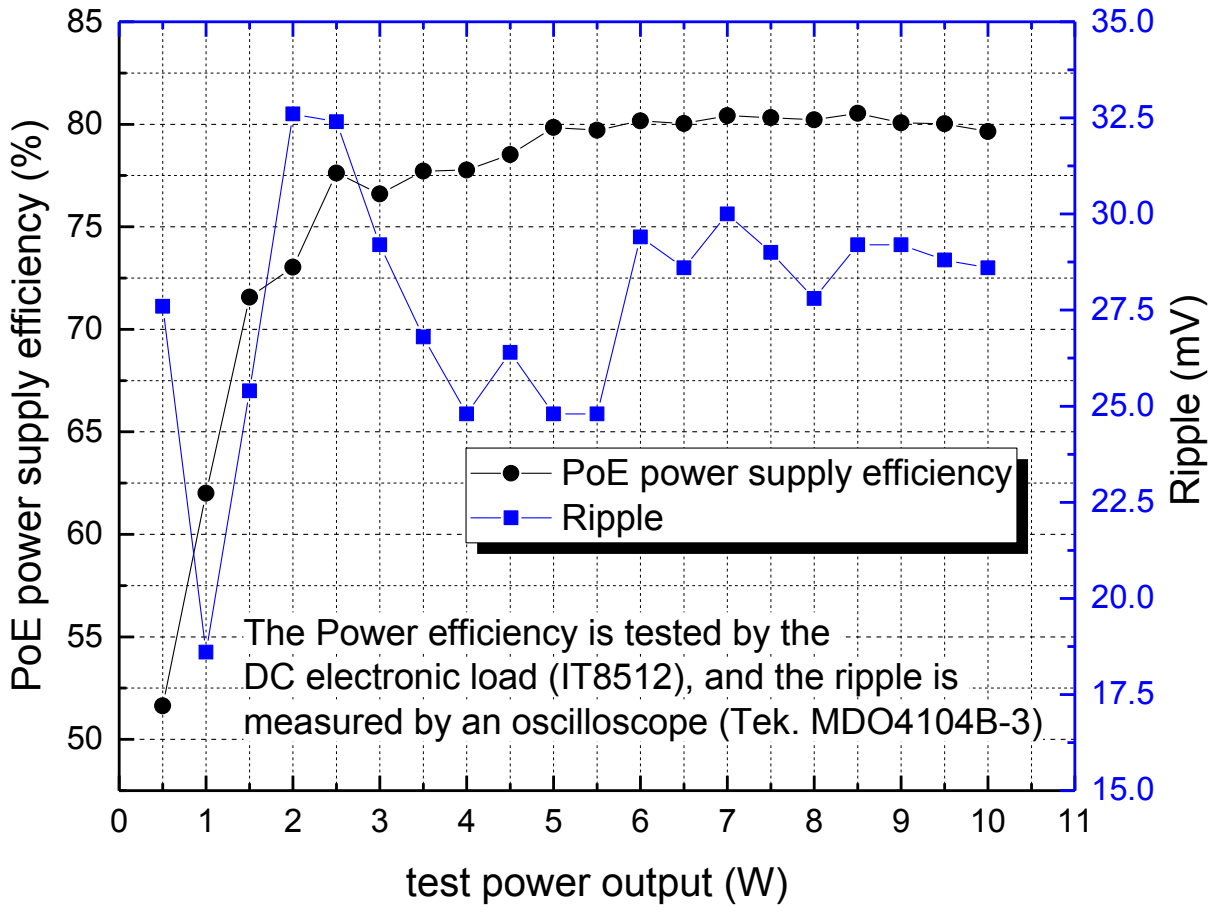

Fig. 5. The efficiency of the PoE PD and its output ripple.

The board of user interface is the front board for user interaction. A standalone microcontroller MKL05Z32VLF4 from NXP Semiconductors is used for critical slow control, human-machine interface, system thermal control, and calibration pulse generator.

### *B. Firmware design*

ZYNQ is a new all-programmable SoC architecture of a 7-series 28 nm field programmable gate arrays (FPGA) with dual-core processors from Xilinx [14]. The development for the ZYNQ readout module includes two sections. One is the firmware design, which is developed by the hardware description language (HDL) for programmable logic (PL). The other section is the software, which is developed in a high-level language such as C or C++.

Fig. 6 depicts the system architecture. It contains the essential parts of the whole system and describes their connections.

The ZYNQ SoC chip in the ZYNQ readout module is the XC7Z020 - CLG400, whose PL architecture is similar with the Artix-7 FPGA. The total DDR SDRAM memory is 512 MB.

The firmware design includes VCO (Voltage-Controlled Oscillator), ADC, and DAC (Digital-to-Analogue Converter) configurations via serial peripheral interface (SPI), ADC data buffering, trigger, and other logical design.

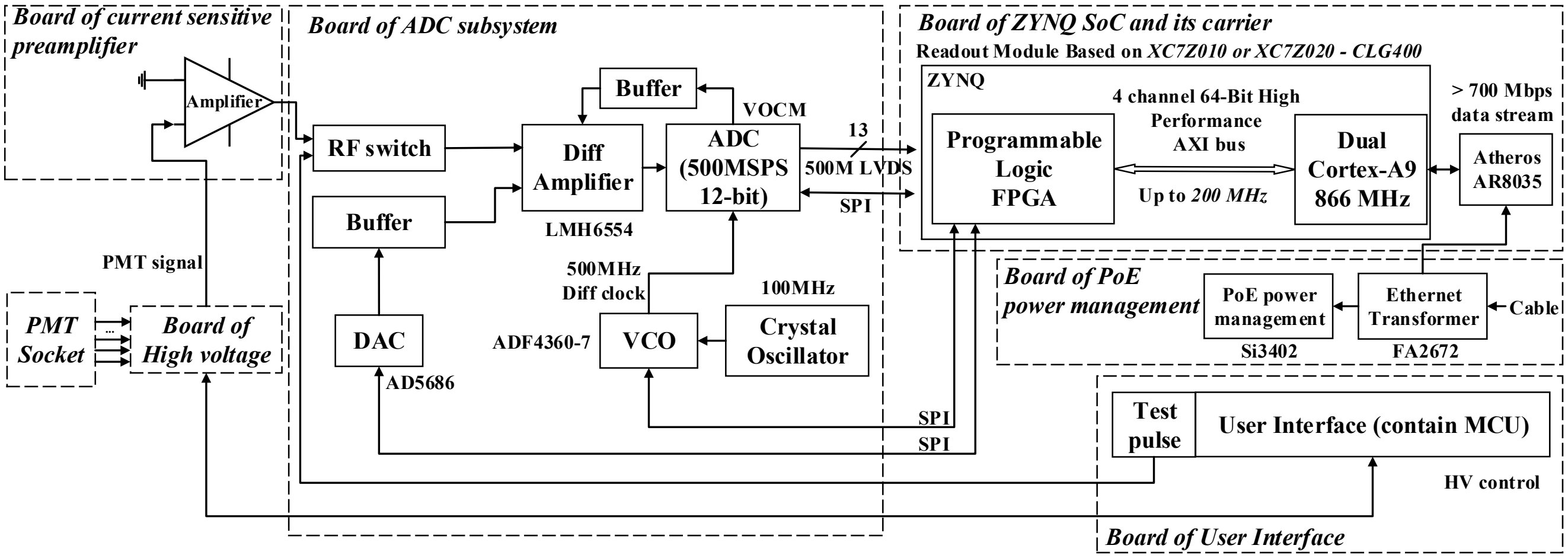

Fig. 6. The system architecture

Fig. 7 illustrates the data processing flow from ADC to a processor system (PS). The IDDR primitive [15] in PL with *SAME_EDGE_PIPELINED mode* is used to deserialize the data stream from ADC with LVDS interface, which will be presented in the next FPGA logic on the same clock edge. Furthermore, a 1:2 FIFO (first in first out) logic deserializes the data and provides a buffer for the ADC clock and global clock from PS. After being stamped with a header, the data stream with 64-bit width is written circularly into the "ring buffer" which consists of dual-port RAM (random access memory).

The global clock 125 MHz is used to read data from the ring buffer and drive the threshold triggering logic, ping-pong logic, and lock RAM, etc. Ping-pong logic is designed to control two BRAMs, so as to shorten the dead time. It is mainly derived from CDMA (central direct memory access), buffer read from DDR memory, and TCP/IP socket transmission.

Table II shows the resource utilization of PL, which is derived from *Vivado 2015.3* IDE (integrated development environment).

TABLE II
THE RESOURCE OF UTILIZATION OF PL.

| Resource | Utilization | Available | Utilization (%) |
|---|---|---|---|
| LUT | 6911 | 53200 | 12.99 |
| LUTRAM | 323 | 17400 | 1.86 |
| FF | 10238 | 106400 | 9.62 |
| BRAM | 36 | 140 | 25.71 |
| IO | 41 | 125 | 32.80 |
| BUFG | 7 | 32 | 21.88 |
| PLL | 1 | 4 | 25.00 |

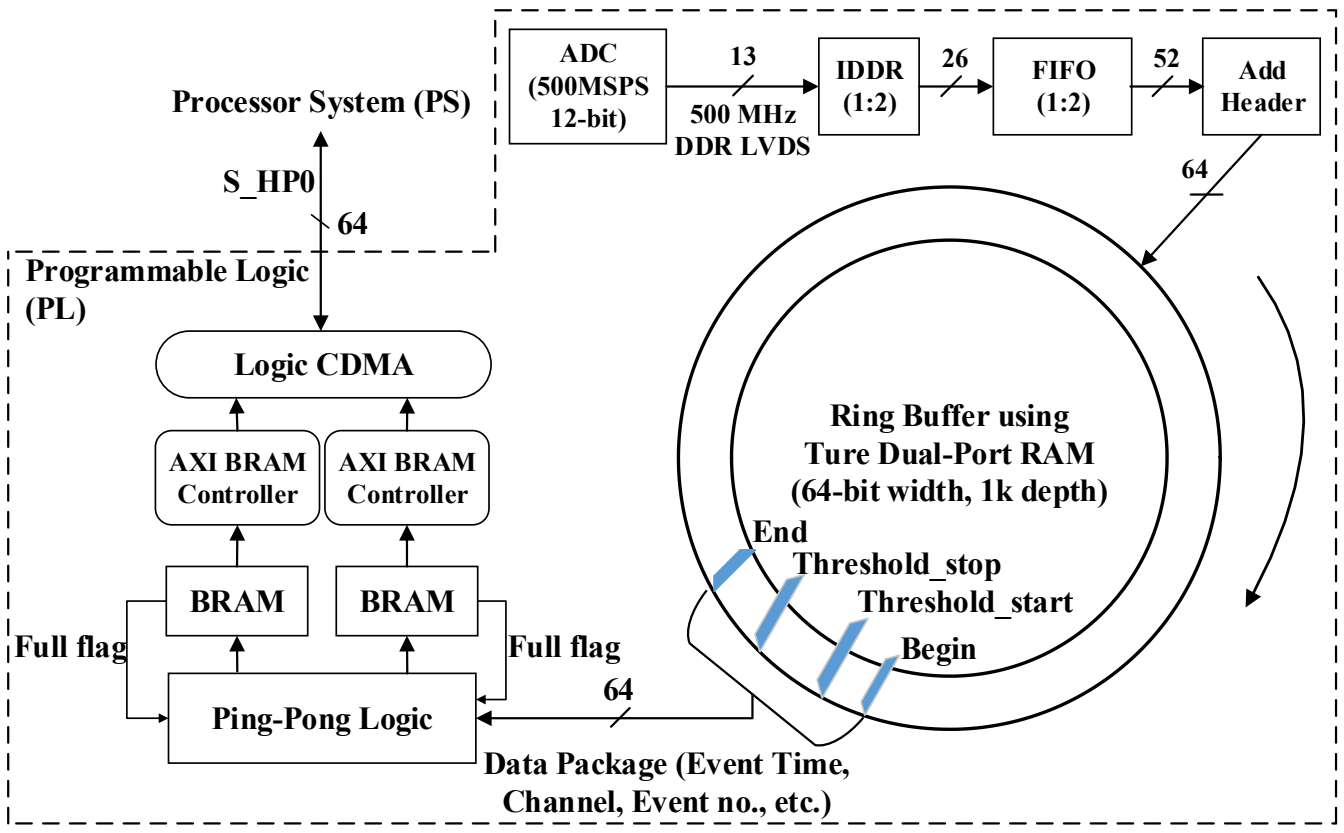

Fig. 7. Data processing flow in PL

## C. Software design

The software design includes the development of the application program (client) in Embedded Linux running on the ZYNQ SoC and a program (remoted server) running on Ubuntu Linux or another platform. File preparations of Embedded Linux can be found in [11]. Fig. 8 and Fig. 9 illustrate the development flow of the Embedded Linux application program and the Ubuntu program, respectively.

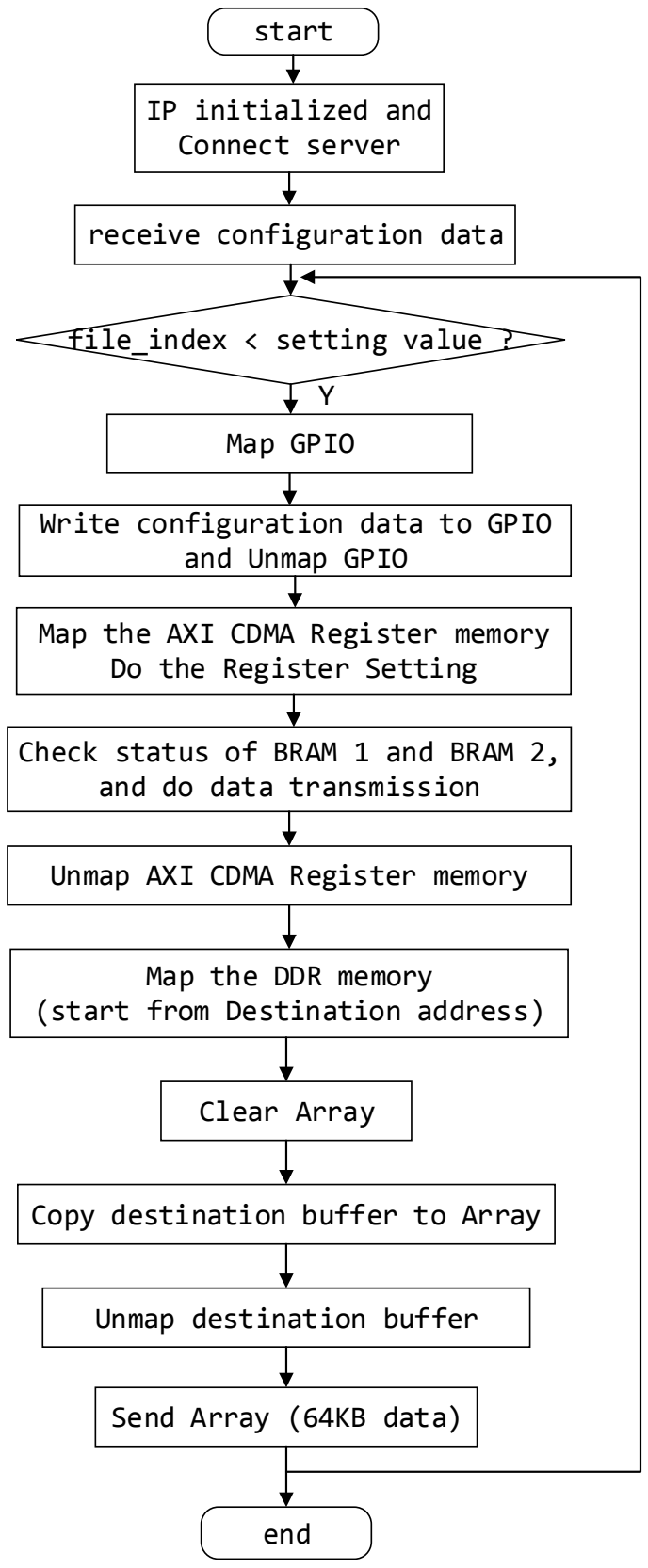

Fig. 8. The development flow of application program (client) on Embedded Linux.

The application program running on Embedded Linux is designed as the client to establish the TCP/IP connection with the remote server. After establishment, the server will read the configuration file, which is from GUI (graphical user interface) settings. A GUI is designed for the user to set parameters including trigger threshold, length of waveform record, DAC tuning output, and pulse polarity.

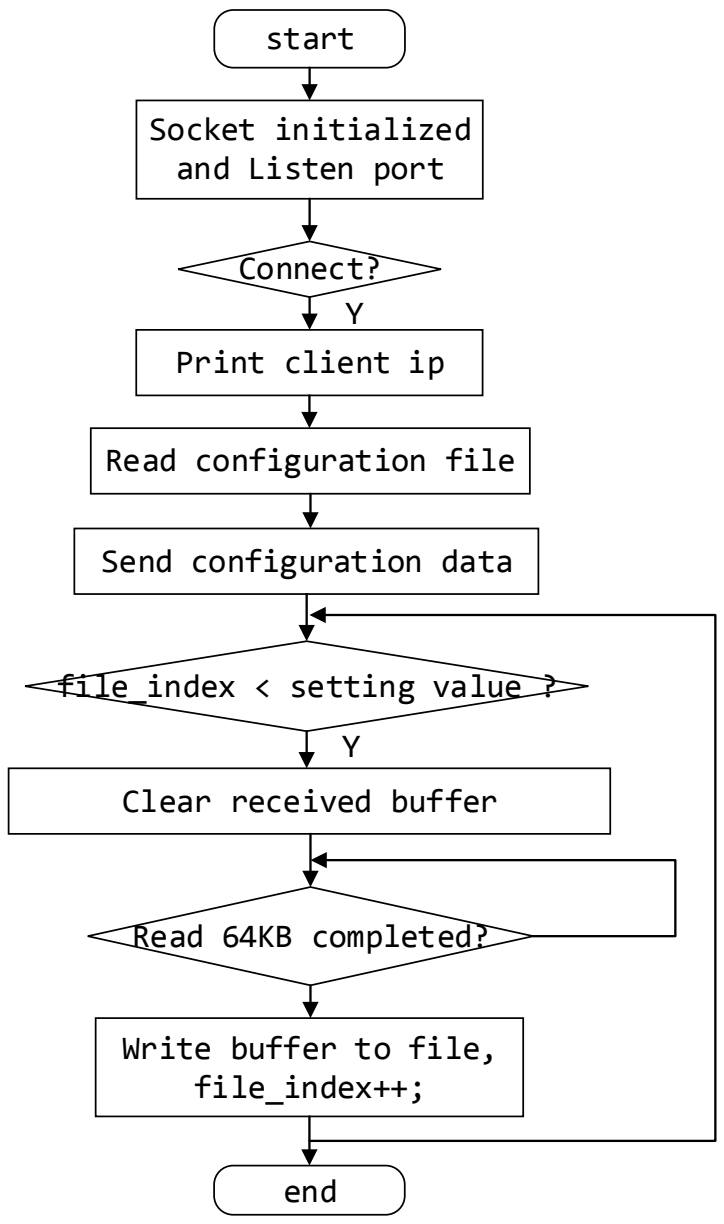

Fig. 9. The development flow of remote program (server) on Ubuntu Linux.

After the client receives the configuration data completely from the server, triggering logic starts. Waveform data over threshold will be buffered into BRAM 1 at first. When BRAM 1 is full (64 KB size), it will be disabled and BRAM 2 will be enabled. Then, CDMA will transfer BRAM 1 data from PL to PS. Finally, the buffer will be sent as a TCP/IP socket.

Because of the discontinuity of the data package, the remote server will receive the whole 64 KB data several times. Then, the received data will be stored in text format. During the experiment, the average CPU usage of Embedded Linux is about 10%.

# III. EXPERIMENT SETUP

## A. Simulation

In order to obtain both fast and thermal neutron signals, moderated and unmoderated $^{252}$Cf source simulations based on Geant4 are done. Fig. 10 presents the geometrical model, and Fig. 11 shows the record of a CLYC scintillator (Φ25.4 mm × 25.4 mm high) with 95% enrichment of $^{6}$Li.

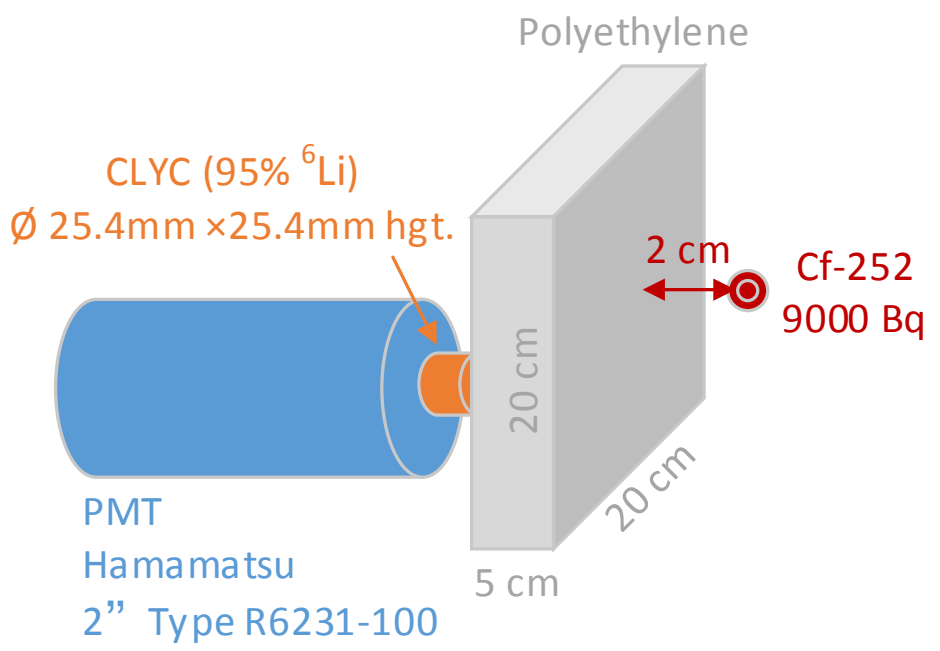

Fig. 10. The geometrical model of simulation.

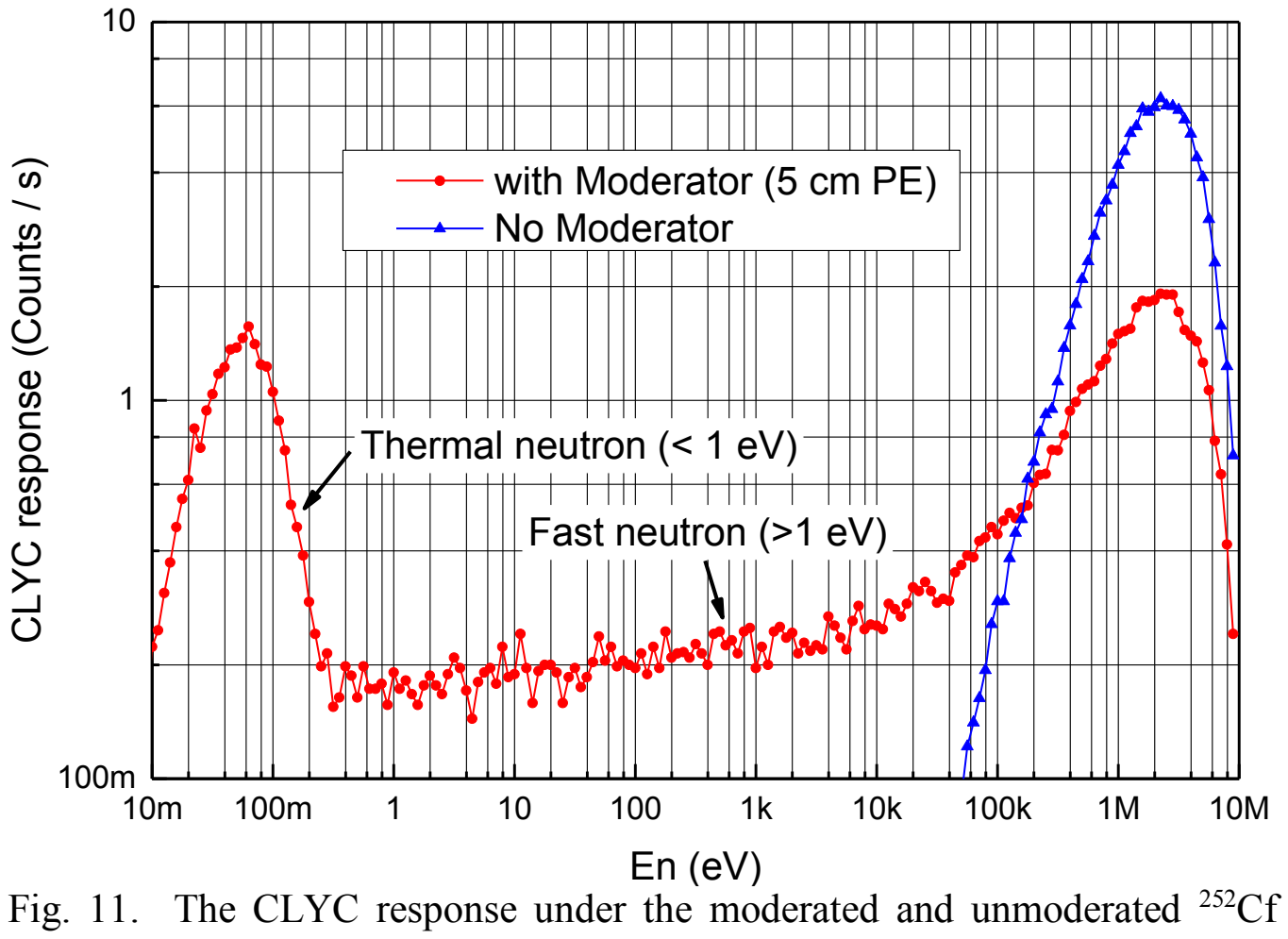

Fig. 11. The CLYC response under the moderated and unmoderated $^{252}Cf$ source. 5 cm polyethylene (PE) is used as the moderator.

From the comparison of the CLYC response, it is known that 5 cm PE can moderate part of fast neutrons, to meet our requirement on neutron-gamma discrimination.

### B. *Source experiments*

Different gamma-ray sources ($^{57}Co$, $^{137}Cs$ and $^{60}Co$) and a moderated neutron source $^{252}Cf$ are used in experiments.

## IV. Results and Analysis

### A. *Pulse shapes and exponential fitting*

The baseline noise is defined as the root mean square (RMS) of the baseline. A threshold of 5 times the baseline noise is set, and 46,000 events are recorded under the moderated $^{252}Cf$ source. Fig. 12 depicts the typical normalized gamma-ray, neutron pulse, and noise shape. The signals are averaged separately to produce the standard pulses. The averaging process reduces the noise observed in the individual signals, making it easier to analyze them. The number of events used to make normalization and its ratio are also presented.

The characteristics of the neutron pulse and gamma-ray pulse exhibit significant differences due to their different scintillation mechanisms. The scintillation properties of CLYC have been studied in [16-18], and four mechanisms are included: direct electron-hole capture ($Ce^{3+}$), binary $V_k$-electron diffusion ($V_k$), self-trapped exciton (STE) emission, and core-valence luminescence (CVL).

The noise shape is the dark count from PMT. It is possible to make pulse shape discrimination (PSD) between the neutron, gamma-ray, and noise. Two integral gates in green are shown. The short integral gate $Q_S$ starts from the wave beginning and has a 100-ns length. The long integral gate $Q_L$ follows and has a 1000-ns length. The *PSD ratio* is defined as equation (3).

Fig. 13 presents the *PSD ratio* distribution. Three kind of pulse shape in Fig. 12 are discriminated only by the *PSD ratio*. The thermal neutrons are detected from $^{6}Li(n,\alpha)t$ reaction, and the fast neutrons from either the $^{35}Cl(n,p)^{35}S$ or the $^{35}Cl(n,\alpha)^{32}P$ reaction. $^{6}Li$ has a thermal neutron capture cross section of 940 barns and a Gamma Equivalent Energy (GEE) of 3.2 MeV approximately. Table III lists the fitting results of decay of mean neutron and gamma-ray pulse. It also compares with Refs. [18] and [16]. The differences may be caused by the different signal sampling rate or the $Ce^{3+}$ doping concentration [18]. Equation (4) is the fitting formula for gamma-ray and Equation (5) is for neutron.

$$PSD\ ratio = \frac{Q_L}{Q_S + Q_L} \quad (3)$$

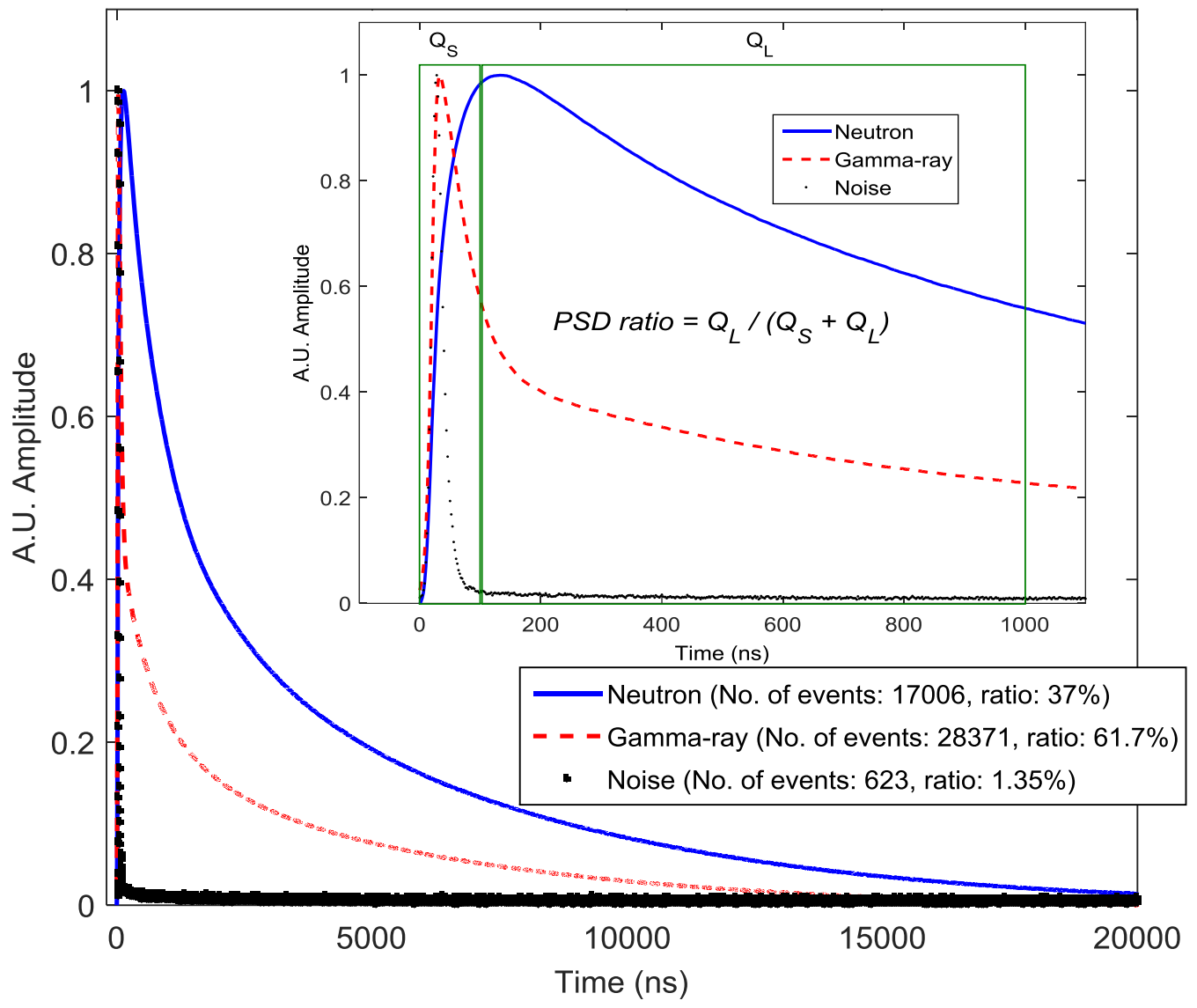

Fig. 12. The typical average gamma-ray, neutron pulse and noise shape emitted by a CLYC with PMT under the moderated $^{252}Cf$ source.

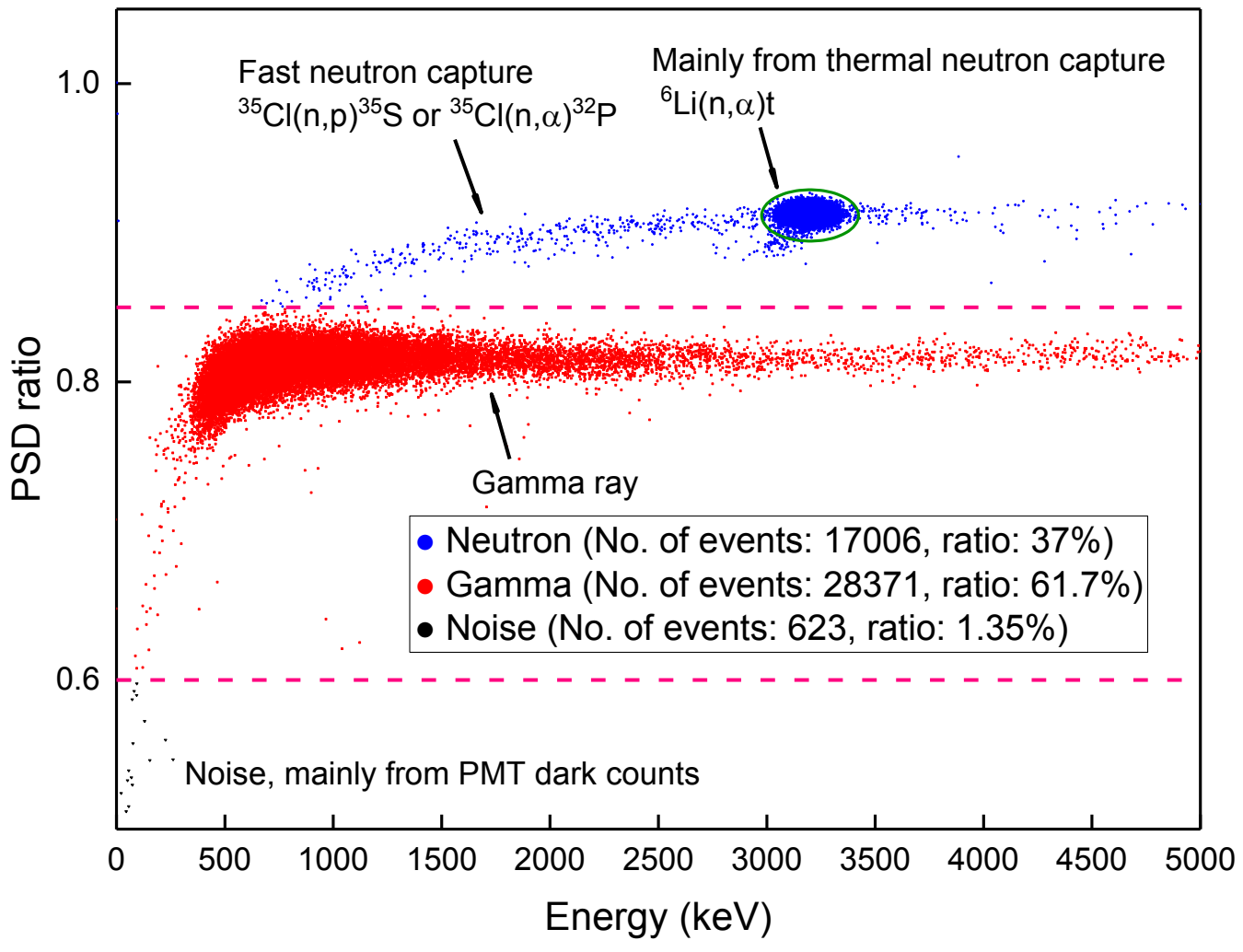

Fig. 13. The PSD ratio distribution.

TABLE III
THE FITTING RESULTS OF DECAY.

| Reference | Particle | CVL (ns) | $Ce^{3+}$ (ns) | $V_k$(ns) | STE(ns) |
|---|---|---|---|---|---|
| This work | Gamma-ray | 49 | 668 | 1141 | 5929 |
| | Neutron | - | 599 | 1339 | 6173 |
| [16] | Gamma-ray | 48 | 280 | 730 | 5240 |
| | Neutron | - | 490 | 1420 | 6300 |
| [18] | Gamma-ray | 2 | 50 | 420 | 3400 |
| | Neutron | - | - | 390 | 1500 |

$$y_{gamma} = 0.53e^{-\frac{t}{49}} + 0.20e^{-\frac{t}{668}} + 0.07e^{-\frac{t}{1141}} + 0.18e^{-\frac{t}{5929}} \quad (4)$$

$$y_{neutron} = 0.40e^{-\frac{t}{599}} + 0.21e^{-\frac{t}{1339}} + 0.41e^{-\frac{t}{6173}} \quad (5)$$

Fig. 14 describes the contribution curves of different mechanisms for the neutron and gamma-ray.

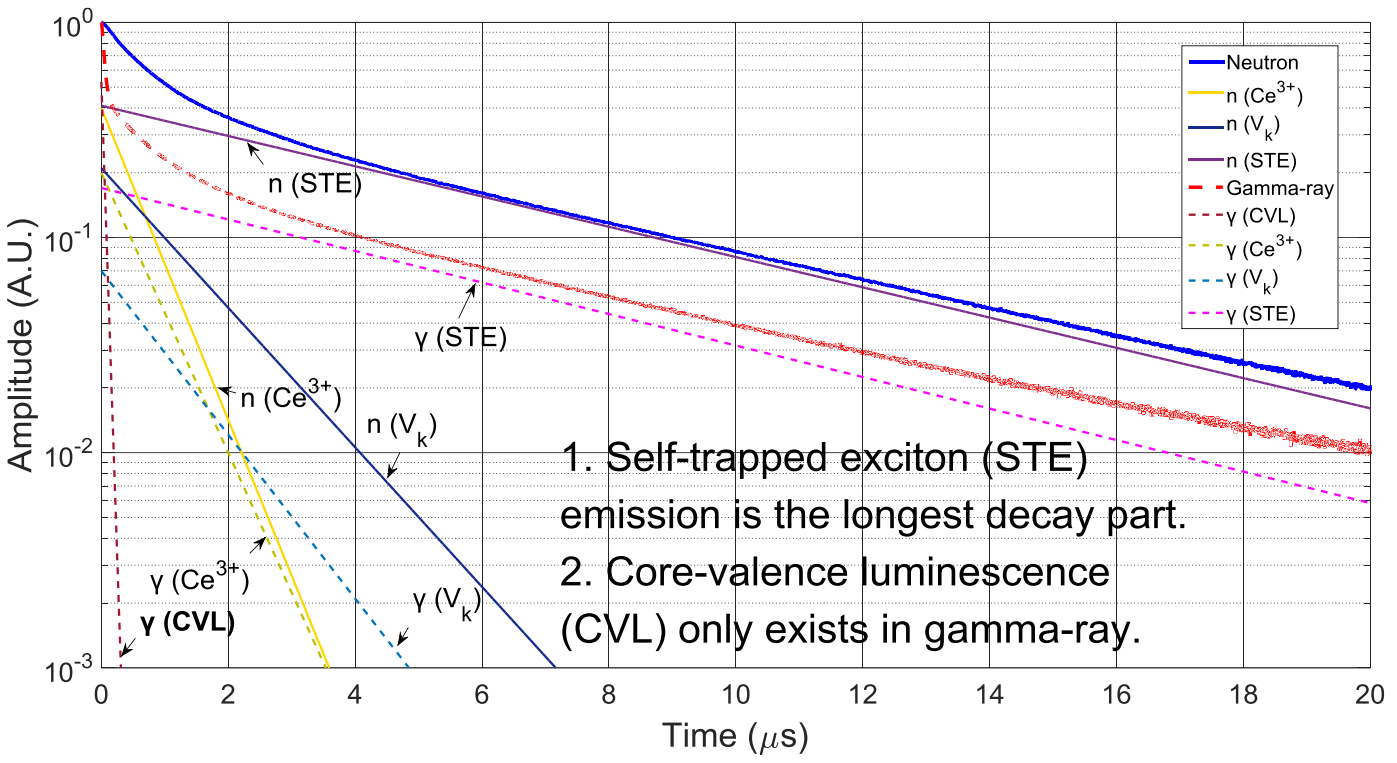

Fig. 14. The contribution curves of different mechanisms for the neutron and gamma-ray.

## B. *Energy resolution and linearity*

The pulse integral is used to obtain the energy of the neutron and gamma-ray. According to Table III and fig. 14, the maximum decay in this work is STE, which is about 6000 ns. We assume that the energy will be collected completely if the pulse integral is 20 µs (more than 3 times of STE decay).

Fig. 15 depicts the ratio of energy collection at different pulse integrals. The energy resolution becomes better as lengthening the pulse integral. It is known that more than 90% energy will be collected and the energy resolution will be better than 5% at 662 keV when the pulse integral is longer than 12 µs.

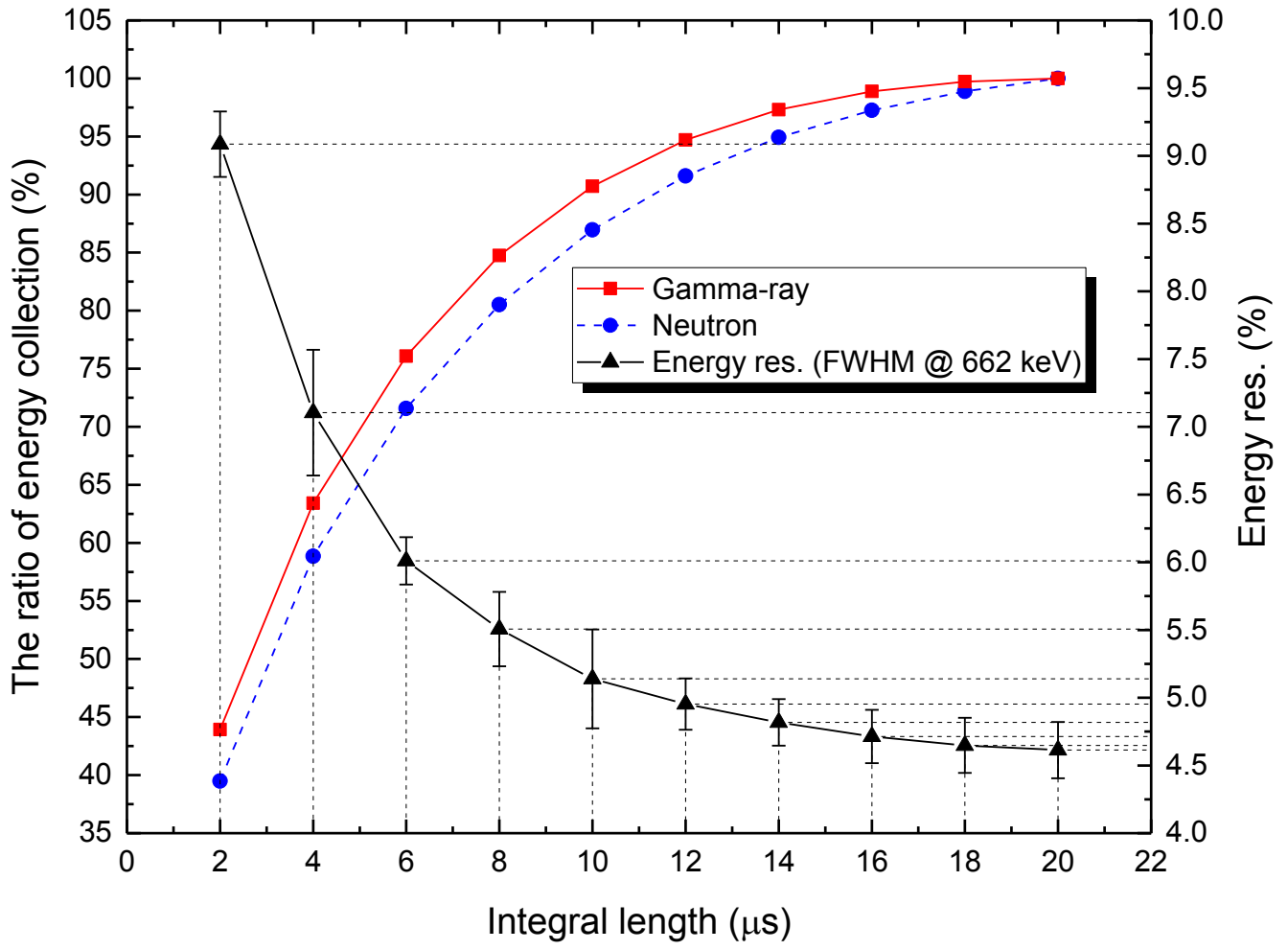

Fig. 15. The ratio of energy collection and the energy resolution at different pulse integrals.

Fig. 16 presents the results of energy resolution in different source experiments with 20 µs integral time. The energy calibration is done by the linear fitting and the GEE of the thermal neutron is estimated to be ~3180 keV. The correlation coefficient $R^2$ is better than 0.9999.

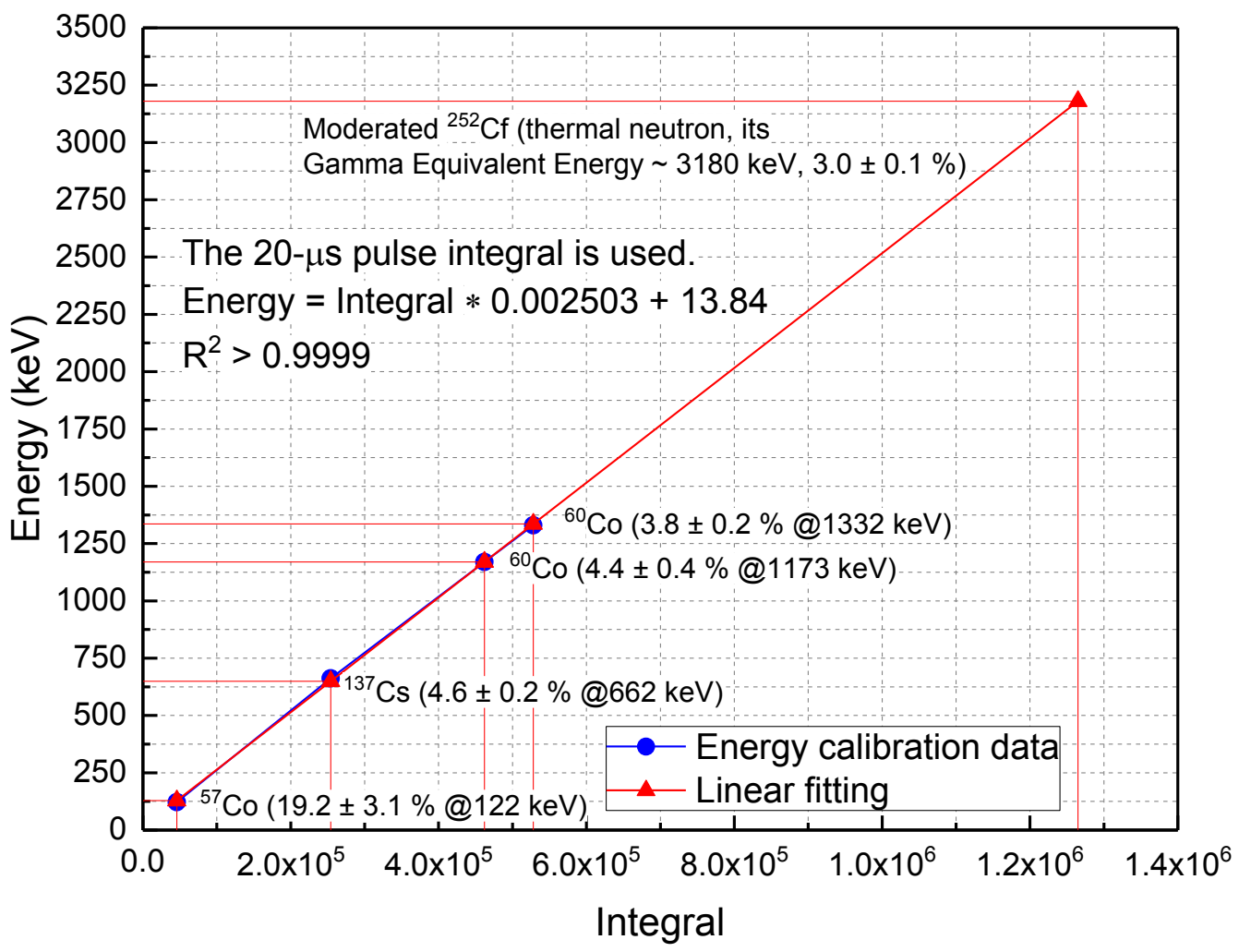

Fig. 16. The energy calibration by the linear fitting.

## C. *Figure of merit*

The whole pulse of the neutron and gamma-ray is used to obtain the figure of merit (FoM). It is calculated by equation (6), where $\mu$ and *FWHM* are the peak location and full width at half maximum of the Gaussian fitting, respectively. Fig. 17 depicts the *PSD ratio* distribution and Gaussian fitting curves in one dimension. The FoM is 3.00 ± 0.03.

$$FoM = \frac{\mu_{neutron} - \mu_{gamma}}{FWHM_{neutron} + FWHM_{gamma}} \quad (6)$$

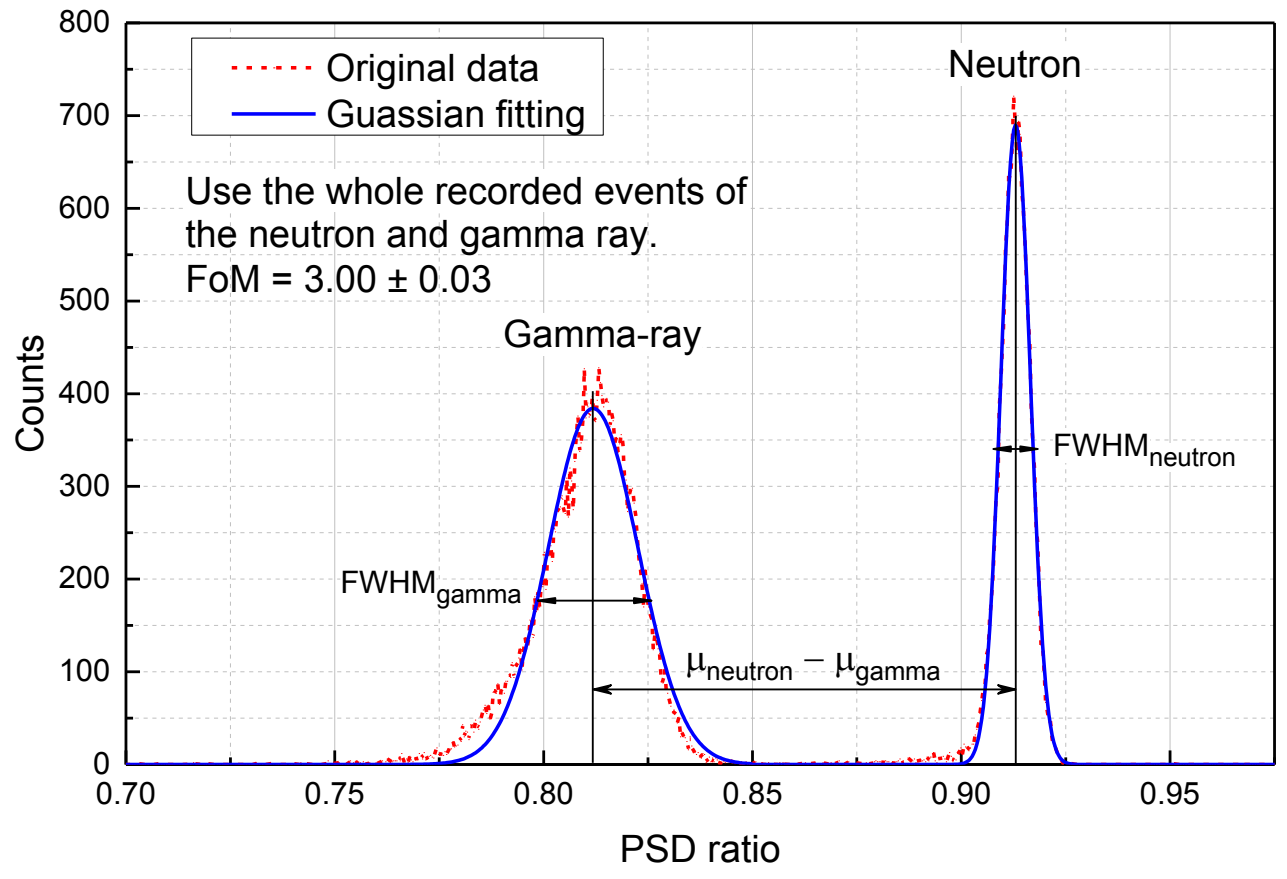

Fig. 17. The one-dimensional PSD ratio distribution and Gaussian fitting curves.

# V. Summary and Future Work

## A. *Summary*

This integrated digitizer base is designed for various applications with PMTs coupled with different detectors. A single category-5 cable is used for power supply and data transmission. The power consumption of the total system is ~5.6 W. A 70 MHz bandwidth preamplifier, 500 MSPS 12-bit ADC board, and a readout module based ZYNQ SoC are designed. In firmware design, threshold triggering with a ring buffer is implemented, and two BRAMs controlled by ping-pong logic are deployed. In software design, the application program (client) on Embedded Linux and a program (remote server) running on Ubuntu Linux are designed for real-time data transmission and storage.

The integrated digitizer PMT base is tested with a CLYC scintillator. The CLYC responses on the moderated and unmoderated $^{252}$Cf source are simulated. Gamma-ray sources and a moderated neutron source $^{252}$Cf are employed in experiments.

The PSD method is used to discriminate between the neutron, gamma-ray, and noise. The decay of the pulse shape of the normalized neutron and gamma-ray is fitted by several exponential formulas. We assume that the total energy is collected by a pulse integral of 20 μs, which is more than 3 times the STE (~6 μs). A 12-μs pulse integral can obtain more than 90% of the total energy and the energy resolution is better than 5% at 662 keV. The linearity is better than 0.9999 from 122 keV to ~3.2 MeV. For the whole pulse of the neutron and gamma-ray, the PSD FoM is ~3.0 using the long integral gate divided by the sum of the long and the short integral gate.

### B. *Future work*

In the future, ADCs with different sampling speeds and vertical resolution will be used in this integrated digitizer base. The influences due to sampling speeds, resolution of ADC, the integral length on PSD FoM, and energy resolution will be analyzed.

In addition, this integrated digitizer base will also be interfaced with the liquid scintillator [19] with the dual R1250 PMTs from Hamamatsu. It is a neutron detector deployed in CJPL (China Jinping Underground Laboratory) for neutron background measurement.

## Acknowledgement

This work is supported by the National Key Research and Development Program of China (2017YFA0402202).

We would like to thank those who collaborated on the CDEX, and also thank Professor Hao Ma, Zhi Deng, Yinong Liu, Zhi Zeng, and Yulan Li for their support and various discussions over the years at the Tsinghua University DEP (Department of Engineering Physics).

We are grateful for the expertise and patient help of Yu Xue, Wenping Xue, and Jianfeng Zhang, technical engineers in the electronics workshop at DEP.